# Performance Metrics Analysis of Torus Embedded Hypercube Interconnection Network

N. Gopalakrishna Kini
Dept of CSE,
Manipal Institute of Technology
(Manipal University), Manipal, India
ng.kini@manipal.edu

M. Sathish Kumar
School of EECS,
Seoul National University,
South Korea
mskuin@yahoo.com

Mruthyunjaya H.S.
Dept. of E&C,
Manipal Institute of Technology
(Manipal University), Manipal, India
mruthyu_hs@manipal.edu

*Abstract*—**Advantages of hypercube network and torus topology are used to derive an embedded architecture for product network known as torus embedded hypercube scalable interconnection network. This paper analyzes torus embedded hypercube network pertinent to parallel architecture. The network metrics are used to show how good embedded network can be designed for parallel computation. Network parameter analysis and comparison of embedded network with basic networks is presented.**

## I. INTRODUCTION

The interconnection network is an important component in a parallel computer. A good interconnection network is expected to have least number of links, topological network cost and more reliable [1]. The interconnection network must be able to scale up with a few building blocks and with minimum redesign. The hypercube is a network with high connectivity and simple routing but the node degree grows logarithmically with number of vertices making it difficult to build scalable architecture [2], [3], [10]. Torus is a network with constant node degree and is highly scalable architecture but has larger network diameter [2], [4], [9].

The advantages of hypercube and torus network can be superposed on to an embedded architecture [4]-[8] called torus embedded hypercube scalable interconnection network.

## II. ARCHITECTURAL PROPERTIES

Let $l \times m$ be the size of several concurrent torus networks with $l$ number of rows and $m$ number of columns and $N$ being the number of nodes connected in the hypercube, the torus embedded hypercube network can be designed with the size of $(l, m, N)$. Nodes with identical positions in the torus networks will be a group of $N$ number of nodes connected in the hypercube configuration and can be addressed with three components such as row number $i$, column number $j$ of torus and address of node $k$ in hypercube where the addressed node is residing. Hence, a $(l, m, N)$–torus embedded hypercube network will have $l \times m \times N$ number of nodes and a node with address as $(i, j, k)$ where $0 \le i < l$, $0 \le j < m$ and $0 \le k < N$.

The data routing functions of torus embedded hypercube network could be analyzed [6]-[8] as in (1)-(5).

$$T_{h1}(i, j, k) = (\ i, (\ j+1)\ mod\ m, k\ ) \qquad (1)$$
$$T_{h2}(i,j,k) = (\ i, (\ m+j-1)\ mod\ m, k\ ) \qquad (2)$$
$$T_{h3}(i, j, k) = (\ (\ i+1)\ mod\ l, j, k\ ) \qquad (3)$$
$$T_{h4}(i,j,k) = (\ (\ l+i-1)\ mod\ l, j, k\ ) \qquad (4)$$
$$T_{Cd}(k_{n-1}.....k_{d+1}\ k_d\ k_{d-1}.....k_0)$$
$$= (k_{n-1}...k_{d+1}\ \overline{k_d}\ k_{d-1}.....k_0) \qquad (5)$$

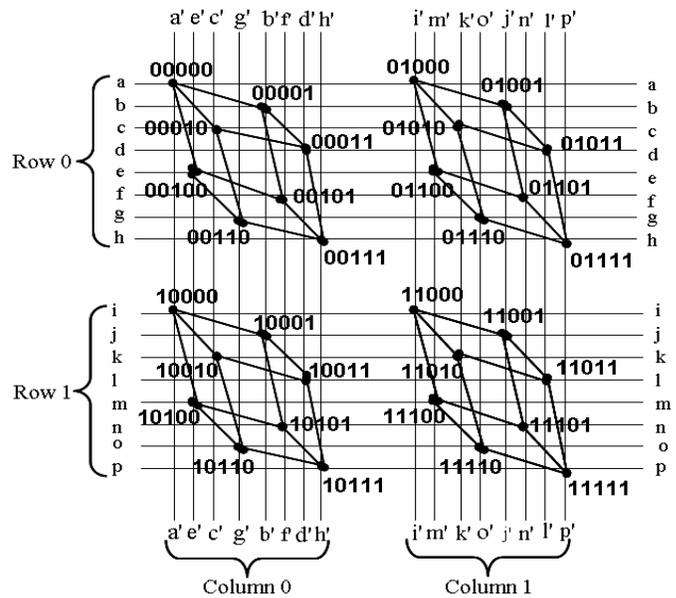

Figure 1.    A (2,2,8)-torus embedded hypercube network

The end to end connections of row and column of each torus are not shown in Figure1 for simplicity. A wraparound connection is done along each row or column if they have same label as a completion of (2, 2, 8)-torus embedded hypercube network.

The proposed network is a highly scalable network. Scalability is achieved in either ways. Firstly, the dimension of the hypercube can be increased by keeping the size of concurrent torus same but increasing the number of concurrent torus accordingly. Secondly, dimension of torus is expanded by keeping the size of the hypercube constant. Scaling up the system using latter method in which expanding the size of torus without affecting the node degree of existing nodes is preferred than the case in former method of hypercube expansion [4], [5].

The total number of links, topological network cost, the scalability and the reliability are the parameters considered in evaluating the performance of this network. The result obtained shows that the torus embedded hypercube favors the scalability of interconnection network.

### III. RESULTS AND DISCUSSION

#### A. Total number of links

For an interconnection network the total number of links is expected to be as low as possible. Because this parameter that reflects link complexity and ultimately the economical cost. Table I shows the number of links with respect to the scaling of the parallel architecture for the basic and embedded networks considered. It is observed that the total number of links for torus embedded hypercube network lies in between n-cube hypercube and torus network. It is to be noted that torus embedded hypercube offers larger number of links than torus network as shown in Figure 2. This is because of the node degree of hypercube that grows logarithmically as the network is scaled up.

#### B. Topological network cost

Network cost is the main parameter for measuring and comparing different topologies. Topological cost of the network depends on the number of links and its diameter. From Table II it is observed that the torus embedded hypercube network has a low network cost. It is also to be observed that the topological cost of ($l$, $m$, 16) - network is more than that of the n-cube hypercube. This is because of the network diameter of the torus network that affects the topological cost of ($l$, $m$, 16) −torus embedded hypercube network. The graphical analysis of network cost of basic and embedded networks is shown in Figure 3.

Even though the torus embedded hypercube network do not reach to the level of prominent features of hypercube or torus network the results obtained shows that it gives better performance with respect to weaknesses of these basic networks.

As far as network scalability of torus embedded hypercube network is concerned, selection of the appropriate scaling up configuration is most important. The configuration is selected in which the dimension of torus is expanded by keeping the size of the hypercube constant and hence the node degree remains constant.

TABLE I. COMPARISON RESULTS OF TOTAL NUMBER OF LINKS OF BASIC AND EMBEDDED NETWORKS

| No. of processors / Network type | 512 | 1024 | 2048 | 4096 | 8192 | 16384 |
|---|---|---|---|---|---|---|
| n-cube Hypercube | 2304 | 5120 | 11264 | 24576 | 53248 | 114688 |
| Torus | 1024 | 2048 | 4096 | 8192 | 16384 | 32768 |
| (16,16,N) – Torus embedded Hypercube | 1280 N=2 | 3072 N=4 | 7168 N=8 | 16384 N=16 | 36864 N=32 | 81920 N=64 |
| (l,m,16) - Torus embedded Hypercube | 2048 | 4096 | 8192 | 16384 | 32768 | 65536 |

TABLE II. COMPARISON RESULTS OF TOPOLOGICAL COST OF BASIC AND EMBEDDED NETWORKS

| No. of processors / Network type | 512 | 1024 | 2048 | 4096 | 8192 | 16384 |
|---|---|---|---|---|---|---|
| n-cube Hypercube | 20736 | 51200 | 123904 | 294912 | 692224 | 1605632 |
| Torus | 22528 | 65536 | 180224 | 524288 | 1474560 | 4194304 |
| (16,16,N) – Torus embedded Hypercube | 21760 N=2 | 55296 N=4 | 136192 N=8 | 327680 N=16 | 774144 N=32 | 1802240 N=64 |
| (l,m,16) - Torus Embedded Hypercube | 20480 | 49152 | 131072 | 327680 | 851968 | 2359296 |

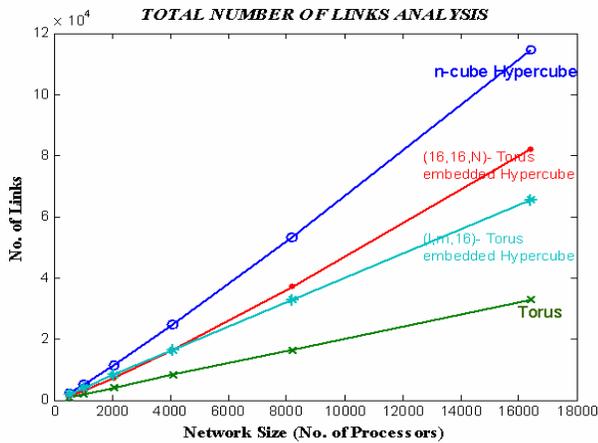

Figure 2. Graphical analysis of number of links of basic and embedded networks

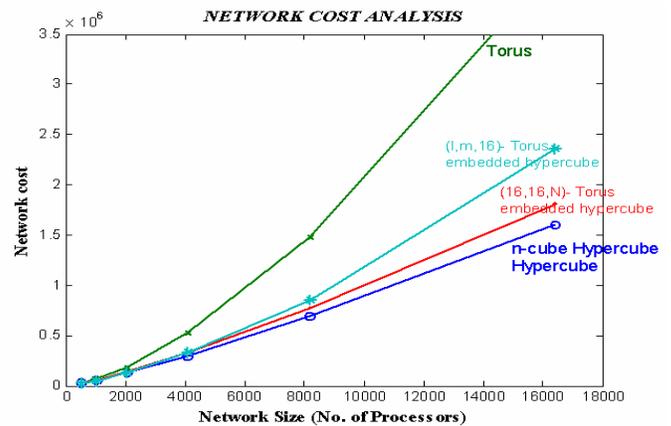

Figure 3. Graphical analysis of network cost of basic and embedded networks

*C. Reliability of the network*

Reliability of a network addresses the probability that a given source-destination pair has at least one fault-free path between them. In the reliability analysis the probability of node or link failure in a scaled up network is considered. An analytical methodology is used in finding the reliability and prediction of availability of interconnection networks. This methodology takes into account the network topology, network size and the routing algorithm used. The reliability analysis presented here is with respect to the failure of the neighboring nodes or links along a routing path.

TABLE III.   RELIABILITY ANALYSIS FOR TORUS EMBEDDED HYPERCUBE INTERCONNECTION NETWORK

| Scaled up networks / Probability of node/link failure | Reliability of network in %age | | | |
|---|---|---|---|---|
| | (4, 4, 8) | (4, 4, 16) | (4, 4, 32) | (4, 4, 64) |
| 1 | 85.7 | 87.5 | 88.9 | 90 |
| 2 | 71.4 | 75 | 77.8 | 80 |
| 3 | 57.1 | 62.5 | 66.7 | 70 |
| 4 | 42.9 | 50 | 55.6 | 60 |
| 5 | 28.6 | 37.5 | 44.4 | 50 |
| 6 | 14.3 | 25 | 33.3 | 40 |
| 7 | 00 | 12.5 | 22.2 | 30 |
| 8 | — | 00 | 11.1 | 20 |
| 9 | — | — | 00 | 10 |

The reliability analysis for the torus embedded hypercube interconnection network is shown in Table III. For a defined network configuration, each and every node possesses equal link complexity. According to the analysis it is observed that the reliability of the torus embedded hypercube interconnection network improves with respect to the scalability of the network. Larger the network better the reliability. Unreliability also gets minimized as the network is scaled up.

## IV.   CONCLUSION

The proposed network is a combination of hypercube and torus network topologies. The analysis results show that torus embedded hypercube interconnection network is highly scalable and configuration of the existing node is not required.

Due to the existence of concurrent multiple torus and hypercubes, this network provides a great architectural support for parallel processing. The growth of the network is more efficient in terms of communication.

Further, an analysis on the reliability of torus embedded hypercube interconnection network has shown that as the interconnection network is scaled up the network will be more reliable and also the unreliability of the interconnection network gets minimized. This is very desirable feature for the interconnection network as the network remains operational for more failure of neighboring nodes or links in parallel computer architecture.